\documentclass[conference]{IEEEtran}
\IEEEoverridecommandlockouts
\usepackage{cite}
\usepackage{amsmath,amssymb,amsfonts}
\usepackage{algorithmic}
\usepackage{graphicx}
\usepackage{textcomp}
\usepackage{xcolor}
\usepackage{mathtools}
\usepackage{commath}
\usepackage{hyperref}
\hypersetup{colorlinks,allcolors=black}
\usepackage[compact]{titlesec}
\titlespacing{\section}{0pt}{2ex}{1ex}
\titlespacing{\subsection}{0pt}{1ex}{0ex}
\titlespacing{\subsubsection}{0pt}{0.5ex}{0ex}
\usepackage{amsmath}
\def\BibTeX{{\rm B\kern-.05em{\sc i\kern-.025em b}\kern-.08em
    T\kern-.1667em\lower.7ex\hbox{E}\kern-.125emX}}
    
\begin{document}
 
\title{Exploiting Partial FDD Reciprocity for Beam Based Pilot Precoding and CSI Feedback in Deep Learning}

\author{Yu-Chien Lin,  Ta-Sung Lee, and Zhi Ding
\thanks{Y.-C Lin is with the Department of Electrical and Computer Engineering,
University of California at Davis, Davis, CA, USA, and
was affiliated with National Yang Ming Chiao Tung University, Taiwan (e-mail: ycmlin@ucdavis.edu).

Z. Ding is with the Department of Electrical and Computer Engineering,
University of California, Davis, CA, USA (e-mail: zding@ucdavis.edu).

T.-S Lee is with the Institute of Communications Engineering, National Yang Ming Chiao Tung University, Taiwan (e-mail: tslee@mail.nctu.edu.tw).}

\thanks{This work is based on materials supported by the National Science Foundation under Grants 2029027 and 2002937 (Lin) and by the Center for Open Intelligent Connectivity under the Featured Areas Research Center Program within the framework of the Higher Education Sprout Project by the Ministry of Education (MOE) of Taiwan, and partially supported by the Ministry of Science and Technology (MOST) of Taiwan under grant MOST 110-2634-F-009-028 and MOST 110-2224-E-A49-001 (Lee and Lin).}}

\maketitle

\begin{abstract}

Massive MIMO systems can 
achieve high spectrum and energy efficiency in downlink (DL)
based on accurate estimate of channel state information (CSI). Existing works have developed learning-based DL CSI estimation that lowers uplink feedback overhead. One often overlooked problem is
the limited number of DL pilots available for CSI estimation.
One proposed solution leverages temporal CSI coherence 
by utilizing past CSI estimates and only sending
CSI-reference symbols (CSI-RS) for partial arrays to preserve CSI
recovery performance. Exploiting CSI correlations,  
FDD channel reciprocity is helpful to base stations with direct
access to uplink CSI.
In this work, we propose a new learning-based feedback 
architecture and a reconfigurable CSI-RS placement
scheme to reduce DL CSI training overhead and to 
improve encoding efficiency of CSI feedback. 
Our results demonstrate superior performance in both
indoor and outdoor scenarios by the proposed framework for CSI recovery at 
substantial reduction of computation power and 
storage requirements at UEs.
\end{abstract}


\section{Introduction}
Multiple-input multiple-output (MIMO) technology and
massive MIMO are vital to 5G and future generations of wireless systems for improvement of
spectrum and energy efficiency. 
The power of massive MIMO hinges on
accurate downlink (DL) channel state information (CSI) at the basestation gNodeB (gNB). Without the benefit of
uplink/downlink channel reciprocity in time-division duplxing (TDD) systems,  gNB of frequency-division duplexing (FDD) systems typically relies on user 
equipment (UE) feedback to acquire DL CSI. 
The extraordinarily 
large number of DL transmit antennas envisioned in
millimeter wave or terahertz bands in future networks \cite{TULVCAN} places a tremendous amount of 
feedback burden on uplink (UL) resources such as
bandwidth and power. As a result, CSI feedback
reduction is crucial to widespread 
deployment of massive MIMO technologies in 
FDD systems. 

Since CSI in most environments has limited delay
spread and can be viewed as sparse, CSI feedback
by UEs can take advantage of such low dimensionality
for CSI feedback compression.  
To extract CSI sparsity for improved feedback efficiency, the work \cite{CsiNet} first proposed a deep autoencoder framework by deploying
encoders and a decoder at UEs and the serving base station, respectively, for CSI compression and recovery. This and other related works have demonstrated
significant performance improvement of CSI recovery
with the use of deep learning autoencoder \cite{ENet, CLNet, CRNet}. 

In addition to autoencoder for direct DL
CSI feedback and recovery, recent works leveraged correlated channel information such as past CSI \cite{CsiNet+,MarkovNet}, 
CSI of nearby UEs \cite{CoCsiNet}, and UL CSI \cite{CQNET, DualNet, DualNet-MP} to improve the recovery of DL CSI
at base stations. Specifically, physical insights considering slow temporal variations of propagation scenarios, similar propagation conditions of similarly located UEs, and  
similarity of
UL/DL radiowave paths reveal significant
temporal, spatial, and spectral CSI correlations 
respectively. More strikingly, UL CSI is generally
available at gNB in existing FDD wireless networks and is
easier to utilize in practice. In addition, 
FDD reciprocity in magnitudes is not only shown from dats
generated by  
CSI models \cite{DualNet} but was also later verified 
in measurement \cite{Zhong2020FDDMM}.
Other related works also considered antenna array geometry
to exploit the UL/DL angular reciprocity
to improve DL CSI estimation in FDD wireless systems \cite{Yacong_reciprocity, Xing_reciprocity}. The work \cite{Yacong_reciprocity} exploited UL/DL angular reciprocity in designing an adaptive dictionary learning for seeking the sparse representation of DL CSIs for feedback. The reciprocity is also utilized for directional training to enhance DL CSI estimation in \cite{Xing_reciprocity}. 

Instead of CSI recovery, a related approach \cite{RH_codebook, 3GPPtypeI, 3GPPtypeII, Wen_codebook} is to exploit FDD reciprocity and angular sparsity to directly determine precoding matrix 
for reducing feedback overhead. The authors \cite{RH_codebook} propose an AoD-adaptive subspace codebook framework for efficiently quantizing and feeding back DL CSI. 
The 5G (NR) supports Type I \cite{3GPPtypeI} and Type II \cite{3GPPtypeII} codebooks corresponding to
low- and high-resolution beams, respectively. The optimum serving beam can be selected by feeding back a predetermined codebook with the largest response between the UE and gNB. Similarly, instead of feeding back predetermined codebook, another
idea in \cite{Wen_codebook} is for UE to feed back 
compressed singular vectors corresponding to the
dominant singular values for precoding matrix optimization. 

Importantly, the estimation accuracy of DL CSI at UEs depends on several factors such as channel fading properties and reference signal (RS) placement. Beyond feedback overhead, the required resource pilot (i.e. CSI-RS) allocation for CSI estimation also grows proportionally with the antenna array size. More resource allocated to CSI-RS would improve
DL CSI estimation accuracy but degrade spectrum efficiency. 
In practical systemsW such as \cite{3GPP}, CSI-RS resources
are sparsely allocated on time-frequency physical resource
grid. 
To our best knowledge, only a few studies \cite{STNet, CANet} considered the sparse CSI-RS availability in designing CSI feedback mechanisms. The deep learning partial CSI feedback framework proposed by \cite{STNet} reduces RS resource 
overhead by leveraging temporal CSI correlation. 
In the work of\cite{CANet}, the gNB optimizes the DL pilot
values (i.e., CSI-RS) based on UL CSI without reducing the
CSI-RS resources. 
However, such implementation
would require dynamic exchange of optimized
pilot values between
the gNB and the UE and is incompatible with the present
use of predefined CSI-RS. 

In this work, we aim to reduce DL CSI-RS overhead 
and the UL feedback overhead while maintaining
DL CSI recovery accuracy at gNB by exploiting 
the available UL CSI. We develop an efficient and reconfigurable deep learning beam based CSI feedback framework by leveraging UL/DL angular reciprocity for FDD wireless systems. Our contributions are summarized as follows:
\begin{itemize}
    \item The framework proposes a beam-space precoding approach
    to exploit the FDD UL/DL reciprocity in beam response magnitudes and generate a low-dimensional representation 
that is easier to recover with fewer antenna ports (APs), leading to lower DL CSI training and UL feedback overhead.
    
    \item The framework reconfigures CSI-RS placement by reducing either pilot resource density or the
number of APs without loss of CSI recovery accuracy. An UL feedback overhead compression module  further reduces UL feedback overhead.

    \item The framework better utilizes FDD reciprocity by not 
    only feeding UL CSI magnitudes as deep learning
    inputs \cite{DualNet}, 
    but also designing a beam-based precoding matrix according to high similarity of UL/DL beam response magnitudes.
    
    \item The reduction of DL CSI training overhead in the framework can significantly lower the computation and storage burdens related to the compression by the
    low cost UEs given the input size reduction of 
    the compression module.
\end{itemize}
We let $(\cdot)^H$, $(\cdot)^T$ denote conjugate transpose and transpose operations, respectively. $(\cdot)^*$ denotes complex conjugate. The $i$-th column of $N\times N$ identity matrix $\mathbf{I}$ is the unit vector $\mathbf{e}_i$.



\section{System Model}
\begin{figure}
\centering
\resizebox{3in}{!}{
\includegraphics*{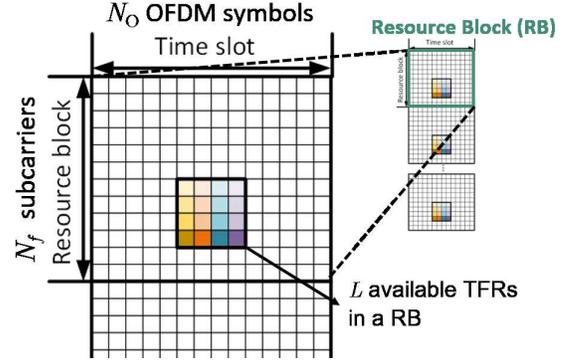}}
\caption{Resource block configuration. \label{fig: RB}}
\end{figure}

We consider a single-cell MIMO FDD link in which
a gNB using a $N_H \times N_V$ uniform planar array (UPA) with $N_b = N_VN_H$ antennas communicates with single antenna UEs. 
Focusing on a specific UE, the DL subband consists of $K$ resource blocks (RBs) for DL CSI-RS and data transmission.
We assume channels within an RB to be
under slow, flat and block fading.
As shown in Fig. \ref{fig: RB}, there are $N_\text{f} \times N_\text{O}$ time-frequency resource elements (REs) in a specific RB ($N_\text{f}$ subcarriers and $N_\text{O}$ OFDM symbols). Since the same processing procedures are applied for every RB, without loss of generality, we only discuss the processing in a single RB in this section.
Given that the gNB assigns $N_b$ REs for DL CSI training for $N_b$ antennas, the received signal vector $\mathbf{y}_\text{DL} \in \mathbb{C}^{N_b\times 1}$ at UE can be expressed as  
\begin{equation}
\setlength{\abovedisplayskip}{4pt}
\setlength{\belowdisplayskip}{4pt}
\mathbf{y}_{\text{DL}}= \mathbf{S}_{\text{DL},N_b}\cdot \mathbf{h}_{\text{DL}} + \mathbf{n}_{\text{DL}}\label{DL signal},
\end{equation}
where $\mathbf{h}_\text{DL} = \text{vec}(\mathbf{H}_\text{DL}) \in \mathbb{C}^{N_b \times 1}$ denotes the DL CSI vector whereas
$\mathbf{S}_{\text{DL},N_b} = \text{diag}(\mathbf{s}_{\text{DL}}) \in \mathbb{C}^{N_b\times N_b}$ denotes the CSI-RS training symbol matrix which is diagonal matrix with diagonal entries of training symbols $s_\text{DL}^{(n)}, n = 1,...,N_b$. $\mathbf{n}_\text{DL} \in \mathbb{C}^{N_b \times 1}$ denotes the additive
noise. $\mathbf{H}_\text{DL} \in \mathbb{C}^{^{N_H \times N_V}}$ denotes the DL CSI matrix before reshaping. From known training symbols in $\mathbf{S}_{\text{DL},N_b}$, the UE
can estimate its DL CSI for feedback to gNB via
\begin{equation}
    \widehat{\mathbf{h}}_\text{DL} = \mathbf{S}_{\text{DL},N_b}^{-1}\cdot \mathbf{y}_\text{DL}. \label{eq: estimated DL CSI}
\end{equation}

\subsection{Beam-Space (BS) Precoding and DL CSI recovery}
Existing wireless systems \cite{3GPP, BSwork1} have applied beamforming/precoding techniques to CSI-RS symbols for 
beam selection, DL CSI estimation, or resistance to attenuation in high frequencies. In this work, we consider DL CSI recovery at 
gNB under beamforming, which serves as CSI performance
baseline. According to \cite{Orthogonal_Beams_Finding}, we can find $N_b$ orthogonal beams to construct an unitary ``orthogonal beam matrix (OBM)'' $\mathbf{B} = [\mathbf{b}^{(1)}\;\mathbf{b}^{(2)}\;...\;\mathbf{b}^{(N_b)}]$. As shown in Fig. \ref{fig:signal flows}.A, 
applying the OBM to the CSI-RS matrix $\mathbf{S}_{\text{DL},N_b}$ in the digital beamforming module, the UE receives 
signals at different REs:
\begin{figure}
\centering
\resizebox{3.4in}{!}{
\includegraphics*{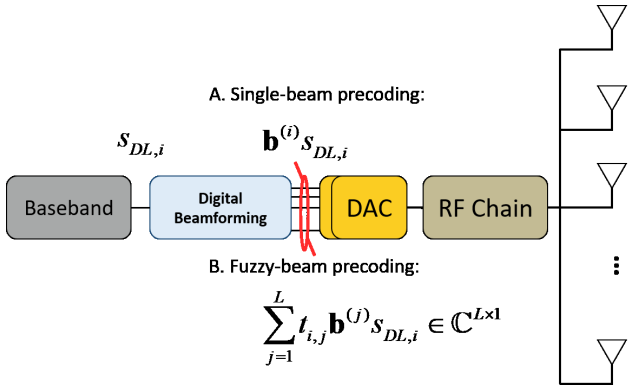}}
\caption{Signal processing flow for beam-space precoding \label{fig:signal flows}}
\end{figure}
\begin{equation}
    \mathbf{y}_\text{DL} = \mathbf{S}_{\text{DL},N_b}\mathbf{B}^T\mathbf{h}_{\text{DL}} + \mathbf{n}_{\text{DL}}.
\end{equation}
From the orthogonality of the OBM, the
DL CSI can reconstructed at the gNB from
the quantized feedback $\bar{\mathbf{g}}_\text{B}=
Q(\mathbf{S}_{\text{DL},N_b}^{-1}\mathbf{y}_\text{DL})$ 
from the UE according to the CSI-RS information 
$\mathbf{s}_{\text{DL}}$ as follows: 
\begin{equation}
    \widehat{\mathbf{h}}_\text{DL} =
     \mathbf{B}^*\bar{\mathbf{g}}_\text{B} =
     \mathbf{B}^*Q(\mathbf{S}_{\text{DL},N_b}^{-1}\mathbf{y}_\text{DL}), \label{CSI recovery with whole beams}
\end{equation}
where $Q(\cdot)$ denotes the encoding process (e.g. quantization).

Given the angular sparsity of DL CSIs, especially for DL CSIs in line-of-sight (LOS) scenarios, the beam space (BS) DL CSI $\mathbf{h}_\text{BS,DL} (=\mathbf{B}^T\mathbf{h}_\text{DL})$ can be assumed as a $L$-sparse vector and thus DL CSI $
\mathbf{h}_\text{DL}$ can be approximated according to the most significant $L \text{ }(L <N_b)$ beams as follows:  
\begin{equation}
    \widehat{\mathbf{h}}_\text{DL} = \mathbf{B}^*_\text{S}\bar{\mathbf{g}}_\text{B,S} \label{CSI recovery with few beams}
\end{equation}
where $\mathbf{B}_\text{S} \in \mathbb{C}^{N_b \times L}$ and $\bar{\mathbf{g}}_\text{B,S} \in \mathbb{C}^{L \times 1}$ respectively denote the significant beam matrix consisting of the steering vectors of the most significant $L$ orthogonal beams, and the corresponding quantized beam responses. Our experiments show that, in propagation channels with low angular spread, the top 1/4 beams approximately contribute to
$90\%$ of DL CSI energy in beam domain. Relying on
$L$ significant beams, the gNB only need to assign $L \text{ }(<N_b)$ REs for CSI-RS in DL to reduce UL feedback.

\begin{figure}
\centering
\resizebox{3in}{!}{
\includegraphics*{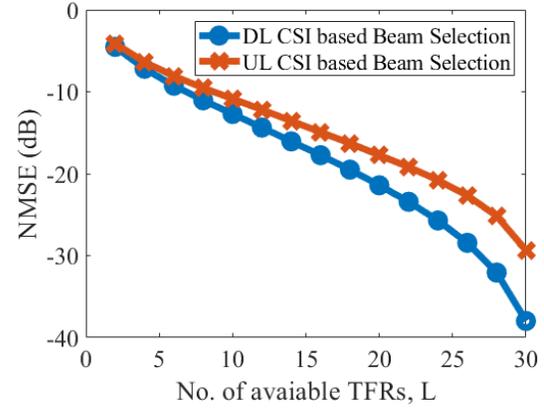}}
\caption{Normalized mean square error (NMSE) of the recovered results obtained by beam selection according to UL/DL CSI magnitudes. (This experiment is based on simulated outdoor UMa channels generated by QuadDRiGa channel simulator \cite{QuaDriGa}.)\label{fig:BS_eval}}
\end{figure}

Typically,
the $L$ significant beams could be found through beam training or direction finding \cite{beam_training, BsNet, low-complexity} by
utilizing additional bandwidth and power
resources. Fortunately, the FDD UL/DL reciprocity in 
magnitudes of angular CSI \cite{DualNet}
can help gNB implement this beam selection process
by relying the available UL CSI at gNB. 
The numerical test results of Fig. \ref{fig:BS_eval} 
illustrate the recovery performance of DL CSI by determining precoding matrix $\mathbf{B}_\text{S}$
which consists of the $L$ significant beams selected according to CSI magnitudes in UL and DL beam domains, respectively. 
The modest difference in terms of CSI estimation error demonstrates the high correlation (reciprocity) between CSI magnitudes in UL and DL beam domains. Specifically, the $L$ dominant beams 
of UL and DL channels are highly correlated. 
Good CSI recovery performance requires 
sufficient number of beams $L$ or REs for CSI-RS.

\section{BS Precoding and DL CSI Recovery}
\subsection{Single-beam Precoding and DL CSI Recovery}
As seen from the preliminary results of Fig. \ref{fig:BS_eval}, 
CSI recovery accuracy hinges on the number of available REs (equal to the number of selected beams). Namely,
missing beam responses of the non-selected beams cause performance degradation. 
On the other hand, careful examination of
the DL CSI in beam domain, we note the significant
spatial correlation between vertically and horizontally adjacent beam responses. Equally important is the
fact that UL CSI magnitudes
can help improve DL CSI estimation. 

\begin{figure}
\centering
\resizebox{3.4in}{!}{
\includegraphics*{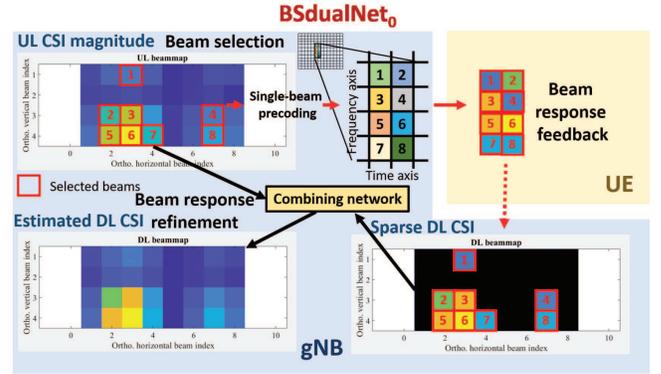}}
\caption{Illustration of $\text{BSdualNet}_0$.\label{fig:BSdualNet0}}
\end{figure}

Taking advantage of these insights, we first develop a heuristic CSI feedback framework, $\textit{BSdualNet}_0$.
As shown in Fig. \ref{fig:BSdualNet0}, the $\text{BSdualNet}_\text{0}$
consists of three phases:
\begin{itemize}
    \item \textit{UL-CSI aided beam selection}: the gNB selects $L$ beams with the largest responses in UL CSI 
by assigning training symbols on $L$ REs for CSI-RS
transmission to UEs. We denote the index set of these beams as $\Omega_\text{B}$. 
    \item \textit{Beam response feedback}: the UE estimates the beam responses for direct encoding and feedback to the gNB.
    \item \textit{Beam response refinement}: the gNB first generates a sparse map filled with the quantized beam responses according to the index set of the selected beams $\Omega_\text{B}$. The sparse map and local
    UL CSI magnitudes form inputs to a deep learning
    network to estimate the missing elements in the 
    sparse map for DL CSI refinement. The
    deep neural network (DNN) generates refined DL beam domain CSI.
\end{itemize}
\subsection{BS Precoding and DL CSI Recovery}
We also develop a BS DL CSI recovery framework which assigns $N_b$ orthogonal beams to $L$ REs ($L<N_b$). Instead of utilizing a single beam for each RE, 
as shown in Fig.~\ref{fig:signal flows}.B, a combination of weighted beams is applied. 
Let us denote an $N_b\times L$ beam merging matrix
\begin{align}
\mathbf{T} = \begin{bmatrix}\mathbf{t}_1\;\;\mathbf{t}_2\;\;...\;\;\mathbf{t}_L\end{bmatrix}, \quad \mathbf{t}_i=
\left[\begin{array}{c} t_{1,i}\\ \vdots\\ t_{N_b,i}
\end{array}\right].
\end{align}

The received signal vector at UE is expressed as
\begin{equation}
\begin{aligned}
    \mathbf{y}_\text{DL} & = \begin{bmatrix}\sum_{i=1}^{N_b}t_{i,1}\mathbf{h}_\text{DL}^{T}\mathbf{b}^{(i)}s_{\text{DL}}^{(1)}\\
    \sum_{i=0}^{N_b-1}t_{i,2}\mathbf{h}_\text{DL}^{T}\mathbf{b}^{(i)}s_{\text{DL}}^{(2)}\\
    \vdots\\
    \sum_{i=0}^{N_b-1}t_{i,L}\mathbf{h}_\text{DL}^{T}\mathbf{b}^{(i)}s_{\text{DL}}^{(L)}
    \end{bmatrix} + \mathbf{n}_\text{DL} \\
    & = \begin{bmatrix}\mathbf{h}_\text{DL}^{T}\mathbf{B}\mathbf{t}_1s_{\text{DL}}^{(1)}\\
    \mathbf{h}_\text{DL}^{T}\mathbf{B}\mathbf{t}_2s_{\text{DL}}^{(2)}\\
    \vdots\\
    \mathbf{h}_\text{DL}^{T}\mathbf{B}\mathbf{t}_Ls_{\text{DL}}^{(L)}\end{bmatrix} + \mathbf{n}_\text{DL}\\
    & = \mathbf{S}_{\text{DL,}L}\mathbf{T}^T\mathbf{B}^T\mathbf{h}_\text{DL} + \mathbf{n}_\text{DL} = \mathbf{S}_{\text{DL,}L}\mathbf{T}^T\mathbf{h}_\text{BS,DL} + \mathbf{n}_\text{DL},
\end{aligned}
\end{equation}
where $\mathbf{T}$ is used to reduce the required REs and to find a compact representation of DL CSI. $\mathbf{h}_\text{BS,DL}=\mathbf{B}^T\mathbf{h}_\text{DL}$ denotes the DL CSI vector in beam domain. The raw and quantized response vectors of the merged beam responses are denoted by $\mathbf{g}_\text{FB} = \mathbf{S}_{\text{DL},L}^{-1}\mathbf{y}_\text{DL}$ and $\bar{\mathbf{g}}_\text{FB} = Q(\mathbf{g}_\text{FB})$, respectively.

Our goal is to find a beam merging matrix $\mathbf{T} \in \mathbb{C}^{N_b \times L}$ and a mapping function $f_\text{re}$ for recovering the DL CSI based on the quantized feedback vector via the principle of
\begin{equation}
    \begin{aligned}
    &\arg\min_{\mathbf{T},\Omega_\text{re}}
    ||\mathbf{B}^*f_\text{re}(Q(\mathbf{S}_{\text{DL},L}^{-1}\mathbf{y}_\text{DL}))-\mathbf{h}_\text{DL}||_\text{F}^2
    \end{aligned}
\end{equation}
where $\Omega_\text{re}$ denotes the 
deep learning model parameters to be 
optimized. Following this principle, the detailed design and architecture of an 
UL CSI-aided feedback framework for DL CSI
estimation will follow in the next section.

\section{Encoder-Free CSI Feedback with UL CSI Assistance}
In this section, we start with the general architecture of the two proposed frameworks (\textit{BSdualNet}, \textit{BSdualNet-MN}). Both exploit UL/DL reciprocity to design the beam merging matrix $\mathbf{T}$ for dimension reduction but utilize different recovery schemes. Next we introduce detailed model learning objectives and design principle. Note that, unlike the previous learning-based frameworks, DNN encoders are not necessary to be deployed on the UEs, thereby reducing memory and computation
burdn on low cost UEs. Instead, this new
framework lowers the required REs for CSI-RS of DL MIMO channels and reduces UL feedback
overhead.

\subsection{General Architecture}
\begin{figure*}[htb]
\centering
\resizebox{6.5in}{!}{
\includegraphics*{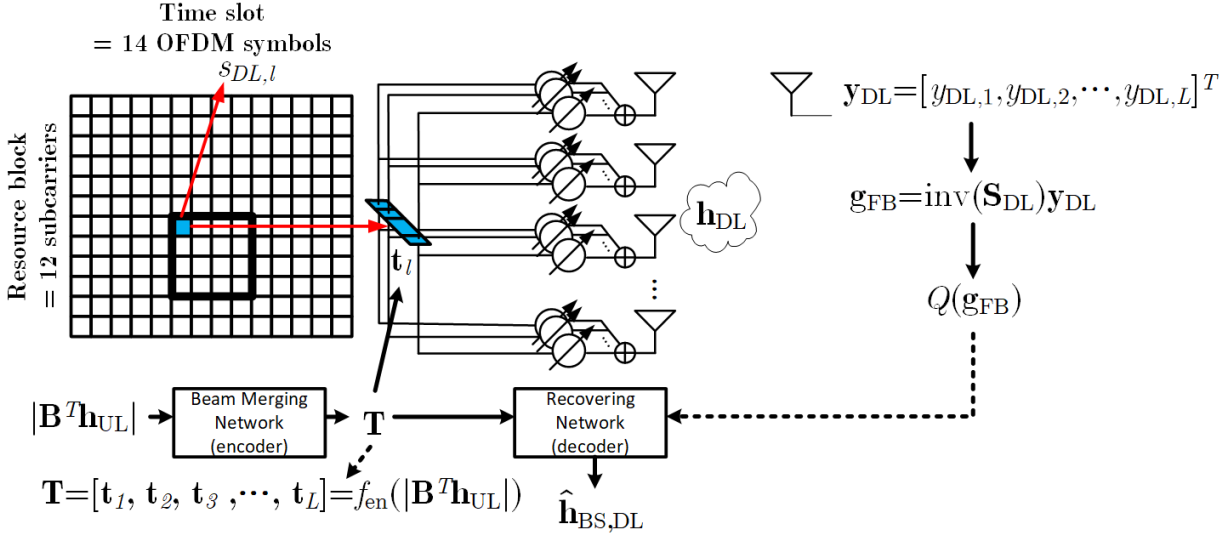}}
\caption{General architecture of the proposed BS CSI feedback framework. (Each of the small grids is a TFR. The region covered by the bold black frame is the designated place for RS replacement. Thus, in this example, the available number of TFRs, $L$, is $16$.)\label{fig:GA}}
\end{figure*}

For simplicity, Fig. \ref{fig:GA} shows the general architecture of the proposed CSI feedback framework for a single-UE, though the same principle applies for multiple UEs.
Consider a wireless communication system with $L$ REs assigned in each RB for CSI-RS placement.
We first design a beam merging matrix $\mathbf{T}$ to match $N_b$ orthogonal beams with different weights to the $L$ REs 
that carry CSI-RS for dimension reduction.
We use a beam merging
network that use UL CSI magnitudes in
beam domain as inputs. Owing to the high correlation between magnitudes of UL and DL CSIs in beam domain, the beam merging network learn to assign suitable weights to  orthogonal beams 
according to the UL CSI magnitudes 
$|\mathbf{B}^T \mathbf{h}_{\rm UL}|$
in BS that are locally available at gNB. 
Next, we apply the beam merging matrix $\mathbf{T}$ to $L$ CSI-RS symbols the $L$ REs. Consequently, 
the effective channels at UEs after CSI estimation would be the weighted sum of 
beam responses as estimate of the
full CSI at downlink. 
Obtaining effective channels, 
the UE simply quantize and feeds back 
the channel information to the gNB. The gNB recovers DL CSI by sending
the quantized feedback and the known beam merging matrix $\mathbf{T}$ into the proposed deep learning decoder network. 

Unlike previous works, our new framework does not require another encoder at UE to 
store and compress full DL CSI. This
is beneficial to UE devices with
limited computation, storage, and/or
power resources. Moreover, we reduce the DL overhead of CSI-RS and provide higher spectrum efficiency. In addition, the linear mapping
matrix $\mathbf{T}$ instead of a general or non-linear mapping function $f: \mathbb{C}^{N_b} \xrightarrow[]{} \mathbb{C}^{L} $ for pilot dimension reduction provides the
advantage of simpler
implementation and easier decoupling of
CSI-RS symbols.

\subsection{BSdualNet}

\begin{figure}
\centering
\resizebox{3.4in}{!}{
\includegraphics*{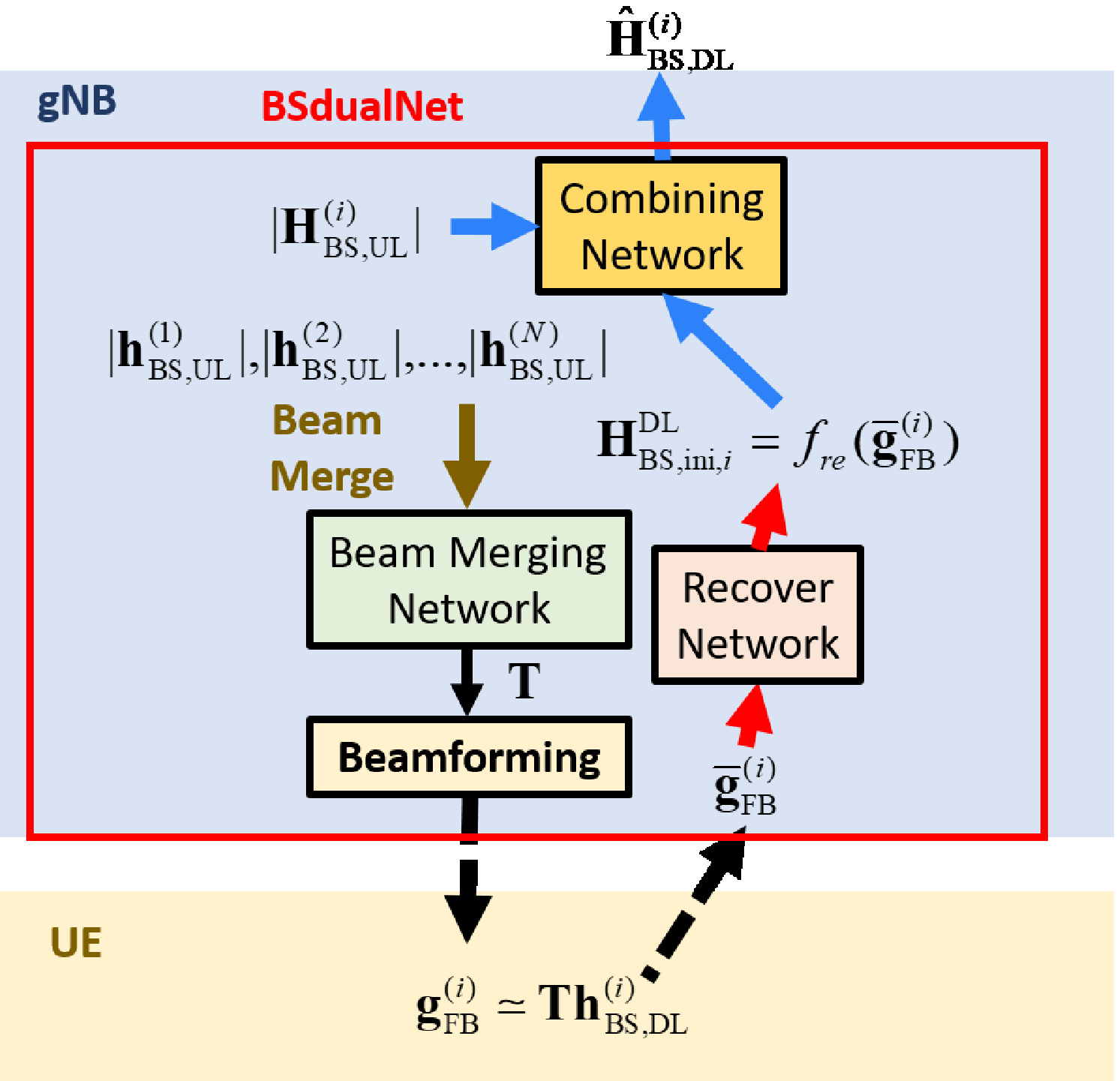}}
\caption{Block Diagram of BSdualNet. \label{fig:BSdualNet_MU}}
\end{figure}

\begin{figure}
\centering
\resizebox{3.4in}{!}{
\includegraphics*{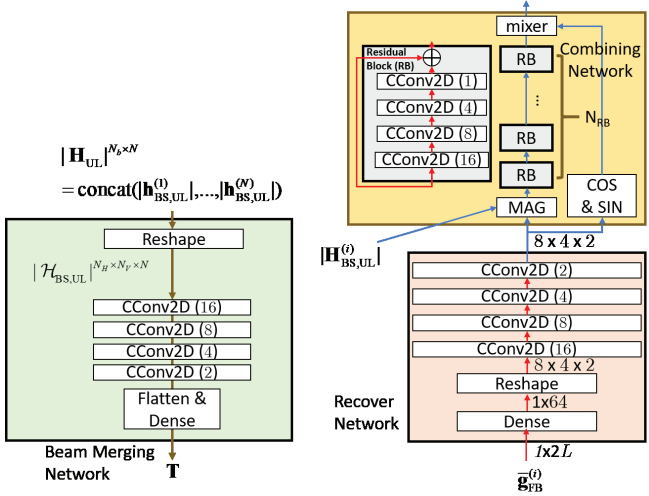}}
\caption{Network design of BSdualNet. \label{fig:BSdualNet_MU_ND}}
\end{figure}

Fig. \ref{fig:BSdualNet_MU} shows the proposed CSI feedback framework, BSdualNet, in multi-user scenarios (i.e., $N$ UEs). As shown in Fig. \ref{fig:BSdualNet_MU_ND}, we aggregate and reshape the magnitudes of BS UL CSIs of each UE into a tensor $|\mathcal{H}_\text{BS,UL}| \in \mathbb{C}^{N_H \times N_V \times N}$, 
which is sent to the beam merging network.
The beam merging deep learning network 
(Fig.~\ref{fig:BSdualNet_MU_ND}) consists of
four $3 \times 3$ circular convolutional layers with 16, 8, 4, and 2 channels, respectively,
to learn the importance of different orthogonal beams according to the spatial structures of UL beam domain CSI magnitudes. 
Given the circular characteristic of BS CSI matrices, we introduce {\em circular convolutional layers} to replace traditional convolution. Subsequently, a fully connected (FC) layer with $2N_bL$ elements is included 
to generate desired dimension after reshaping (Recall that $\mathbf{T}$ is a complex matrix with size of $N_b \times L$). 
After CSI estimation at UEs, the gNB receives the $N$ copies of quantized feedbacks from $N$ UEs and obtains quantized feedbacks $\bar{\mathbf{g}}_{\text{FB}}^{(i)} \in \mathbb{C}^{2L}, i = 1,2,\dots,N$.

Now we focus on the network at gNB. For the $i$-th UE, we forward the received feedback $\bar{\mathbf{g}}_{\text{FB}}^{(i)}$ to a FC layer with $2N_b$ elements. After reshaping 
the feedback data into a matrix of size $N_H \times N_V \times 2$, we use
four $3 \times 3$ circular convolutional layers with $16$, $8$, $4$, and $2$ channels and activation functions to generate initial BS DL CSI estimate $f_\text{re}(\bar{\mathbf{g}}_{\text{FB}}^{(i)})$. 
Next, the gNB forwards the initial BS DL CSI estimate $f_\text{re}(\bar{\mathbf{g}}_{\text{FB}}^{(i)})$ together with the BS UL CSI magnitudes $|\mathbf{H}_\text{BS,UL}^{(i)}|$ to the combining network for final DL CSI estimation.
The combining network uses $N_B$ residual blocks, each block contains the same design of  circular convolutional layers and activation functions as the network for DL CSI recovery.

The BSdualNet is optimized by updating the network parameters $\Theta_\text{bm}$, $\Theta_\text{re}$ and $\Theta_\text{c}$ of non-linear beam merging, recovery, and combining networks $f_\text{bm}$, $f_\text{re}$ and $f_\text{c}$:
\begin{equation}
   \mathop{\arg\min}_{\Theta_\text{bm}, \Theta_\text{re}, \Theta_\text{c}}
   \left\{\sum_{i=0}^{N-1}\norm{\widehat{\mathbf{h}}_{\text{BS,DL}}^{(i)}-\mathbf{h}_{\text{BS,DL}}^{(i)}}^2_\text{F}
   \right\}, \\ \label{loss_magnitude}
\end{equation}
\begin{equation}
    \widehat{\mathbf{h}}_{\text{BS,DL}}^{(i)} = f_\text{c}(f_\text{re}(\bar{\mathbf{g}}_{\text{FB}}^{(i)}), |\mathbf{H}_{\text{BS,UL}}^{(i)}|),
\end{equation}
\begin{equation}
    \bar{\mathbf{g}}^{(i)}_\text{FB} = Q((\mathbf{S}^{(i)}_{\text{DL},L})^{-1}\mathbf{y}^{(i)}_\text{DL}),
\end{equation}
\begin{equation}
\begin{aligned}
    \mathbf{y}_\text{DL}^{(i)} = \mathbf{S}^{(i)}_{\text{DL},L}\mathbf{T}\mathbf{h}^{(i)}_\text{BS,DL} + \mathbf{n}^{(i)}_\text{DL},
\end{aligned}
\end{equation}
\begin{equation}
    \mathbf{T} = f_\text{bm}(|\mathbf{h}_{\text{BS,UL}}^{(1)}|,|\mathbf{h}_{\text{BS,UL}}^{(2)}|,...,|\mathbf{h}_{\text{BS,UL}}^{(N)}|).
\end{equation}
Note that the superscript ${}^{(i)}$ denotes the UE index. $\mathbf{h}_{\text{BS,UL}}^{(i)} = \text{vec}(\mathbf{H}_{\text{BS,UL}}^{(i)}) \in \mathbb{C}^{N_HN_V}$ and $\mathbf{H}_{\text{BS,UL}}^{(i)} \in \mathbb{C}^{N_H \times N_V}$ denote the vectorized and original UL CSI in beam domain at the $i$-th UE.

\subsection{BSdualNet-MN}
In BSdualNet, the beam merging network provides a beam merging matrix $\mathbf{T}$ to generate an efficient representation of the convoluted responses of all orthogonal beams. Although $\mathbf{T}$  is optimized for the ease of
decoupling individual beam responses, the decoder remains a blackbox such that the information within $\mathbf{T}$  may not be fully exploited due to its indirect use.
In this section, we would redesign the decoder by directly using the beam merging matrix 
$\mathbf{T}$ to achieve better architectural interpretability and performance improvement.

\begin{figure}
\centering
\resizebox{3.4in}{!}{
\includegraphics*{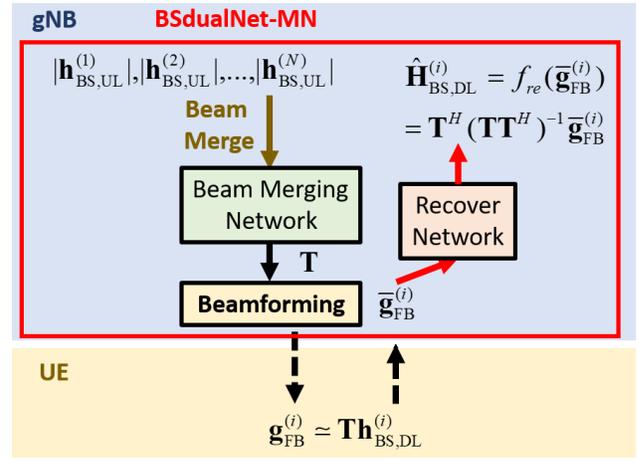}}
\caption{Block Diagram of BSdualNet-MN.\label{fig:BSdualNet_MU_MN}}
\end{figure}

Unlike the previous works that split the
deployment of CSI encoder and decoder at UEs and gNB, respectively, our gNB knows the exact 
encoding and decoding processes in our framework. Thus, we can exploit the locally known beam merging matrix $\mathbf{T}$ to decode the feedback more efficiently. 
To this end, we reformulate the problem of DL CSI recovery for
$\widehat{\mathbf{h}}_{\text{BS,DL}}^{(i)},i=0,...,N-1$ by seeking a minimum-norm solution 
to an under-determined linear system
\[\mathbf{y}_{\text{DL}}^{(i)}=\mathbf{T}^T\mathbf{h}_{\text{BS,DL}}^{(i)}+ \mathbf{n}^{(i)}_\text{DL},i=0,...,N-1.
\]
As 
seen from Fig. \ref{fig:BSdualNet_MU_MN}, the output of the recovery network can be expressed as follows: 
\begin{equation}
\begin{aligned}
    f_\text{re}(\widetilde{\mathbf{g}}_{\text{FB},i}^{(i)}) & = \mathbf{T}^H(\mathbf{T}\mathbf{T}^H)^{-1}\widetilde{\mathbf{g}}_{\text{FB}}^{(i)},
\end{aligned}\label{eq: MN}
\end{equation}
Clearly, the minimum norm solution depends on matrix $\mathbf{T}$. Assuming
perfect quantization and zero noise, we can 
approximate the decoder\footnote{See Appendix} of Eq. (\ref{eq: MN}) as
\begin{equation}
\begin{aligned}
    f_\text{re}(\widetilde{\mathbf{g}}_{\text{FB}}^{(i)}) & \approx \mathbf{T}^H(\mathbf{T}\mathbf{T}^H)^{-1}\mathbf{T}\mathbf{h}_{\text{BS,DL}}^{(i)},\\
    & = \underbrace{\sum_{i=1}^{L}\mathbf{v}_i\mathbf{v}_i^H}_{\widetilde{\mathbf{I}}}\mathbf{h}_{\text{BS,DL}}^{(i)}= \widetilde{\mathbf{I}}\cdot \mathbf{h}_{\text{BS,DL}}^{(i)}, \label{eq: I}
\end{aligned}
\end{equation}
where $\mathbf{v}_i, i = 1,2,\dots,N_b$ are 
right singular vectors of $\mathbf{T}$. Since  $Trace(\widetilde{\mathbf{I}})=L$, $\mathbf{h}_{\text{BS,DL}}^{(i)}$ cannot be fully recovered by only relying on the diagonal entries of $\widetilde{\mathbf{I}}$. If strong spatial correlation exists in the beam domain, we will need a recovery matrix $\widetilde{\mathbf{I}}$ with larger off-diagonal entries, representing the correlation between beams. Given the FDD UL/DL reciprocity in beam domain, by capturing the correlation between adjacent beam response magnitudes of UL CSI, it would be more reasonable to define a
merging matrix $\mathbf{T}$ which contains well-behaved right singular vectors such that 
$\sum_{i=0}^{N-1}||\widetilde{\mathbf{I}}\mathbf{h}_{\text{BS,DL}}^{(i)} - \mathbf{h}_{\text{BS,DL}}^{(i)}||_\text{F}^2$ 
can be minimized.

With the same design of the beam merging network in BSdualNet, the recovery network in BSdualNet-MN simply includes a series of matrix products. Thus, BSdualNet-MN is not only more interpretable, its computational complexity and required model memory are also lower.

\section{UL CSI Aided Beam Based Precoding and a Reconfigurable CSI Feedback Frameworks}
Generally, the aforementioned methods perform better with high sparsity CSI in beam domain. Yet, such spatial sparsity may not hold for CSI of 
every propagation channels. For example, indoor propagation channels tend to exhibit rich multi-paths with high angular spreads. This could lessen spatial sparsity and degrade recovery accuracy of DL CSI. 
Interestingly, however, such channels are 
alternatively characterized by large
coherence bandwidth because of the
dominance of 
low-delay paths dominate\cite{CSImeasurement}.
This means that for such channels, it is not necessary to have high CSI-RS density in frequency domain.

In this section, a reconfigurable CSI feedback framework will be described as a more flexible solution to reduce the number of pilots by selecting frequency reduction (FR) and beam reduction (BR) ratios. Instead of regarding feedback of each RB independently,
as discussed in the signal model of Section II, we
exploit the large coherence bandwidth 
and consider a joint UL feedback for a total of $K$ RBs. 
By leveraging spectral coherence, we can further reduce the UL feedback overhead by applying an autoencoder network. 
In what follows, we elaborate on the reconfiguration of CSI-RS placement and the design of a learning-based CSI feedback framework, BSdualNet-FR.

\subsection{Frequency Resource Reconfiguration}
\begin{figure}
\centering
\resizebox{3in}{!}{
\includegraphics*{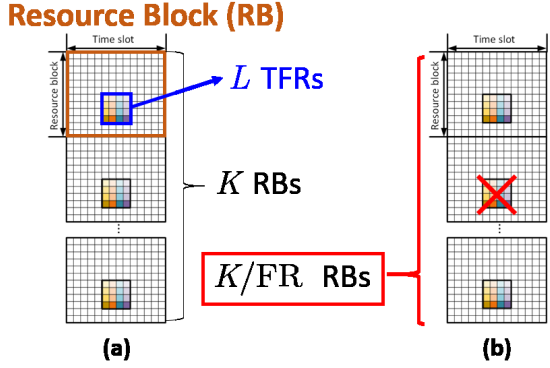}}
\caption{Illustration of pilot number reduction. (Note that the color grids represent the designated TFRs in one of the pilot placement configurations defined in 5G specification \cite{3GPP}.)\label{fig:pilot reduction}}
\end{figure}

\begin{figure}
\centering
\resizebox{3.4in}{!}{
\includegraphics*{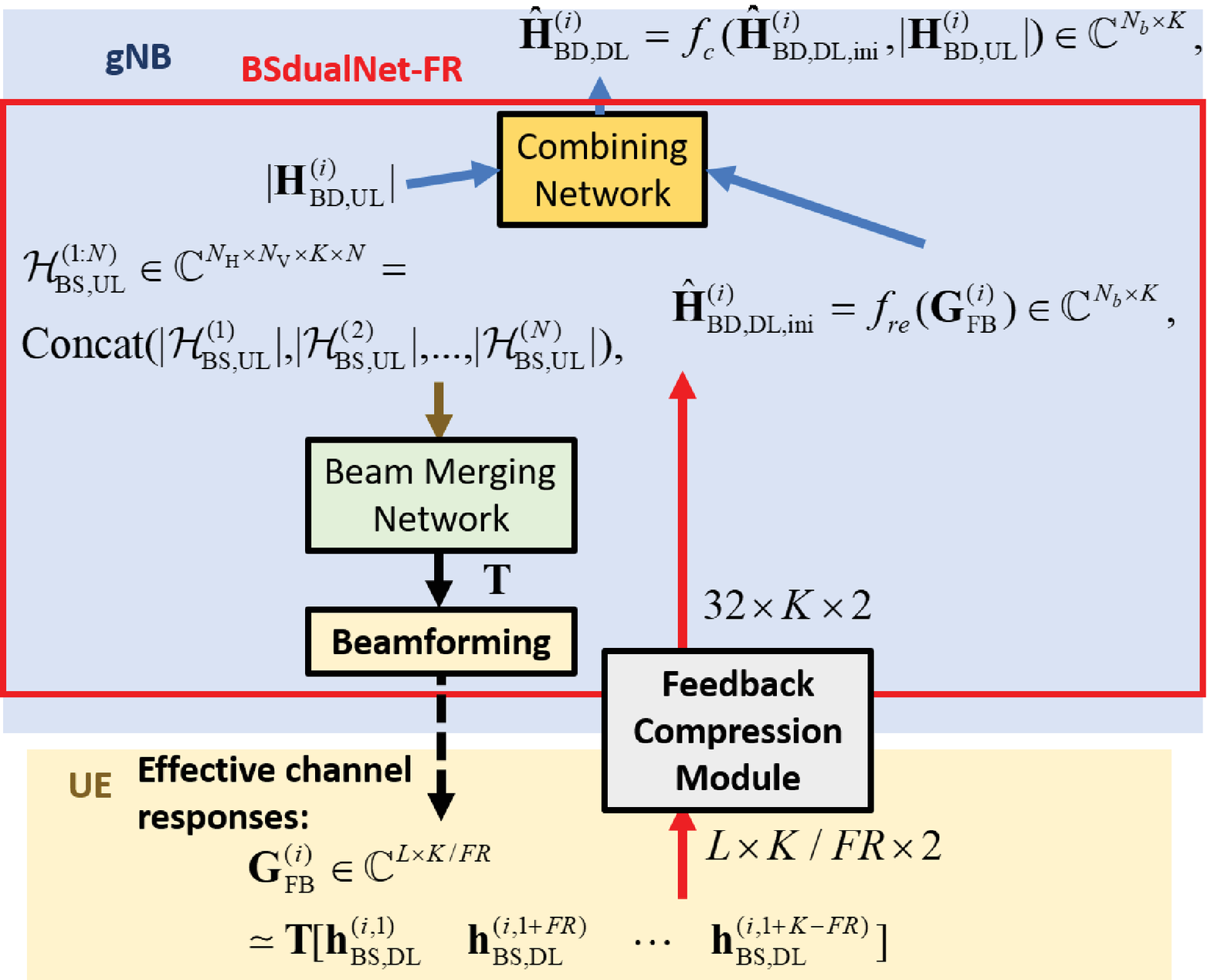}}
\caption{Block Diagram of BSdualNet-FR. \label{fig:BSdualNet_FR}}
\end{figure}

\begin{figure}
\centering
\resizebox{3.4in}{!}{
\includegraphics*{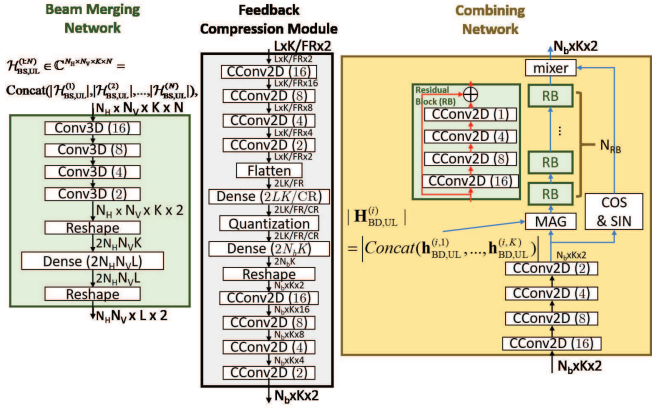}}
\caption{Network design of BSdualNet-FR. \label{fig:BSdualNet_FR_ND}}
\end{figure}

In modern wireless protocols, there are designated resource regions for CSI-RS placement \cite{3GPP}. Compatible with existing RS configurations, we can reduce the CSI-RS placement density along the frequency domain by a frequency
reduction factor $\textit{FR}$ by placing pilots only at RB indices $k = 1, 1+FR, 1+2FR,..., 1+(K-1)FR$ as shown in Fig.~\ref{fig:pilot reduction}. 
We can also further reduce the required REs by a beam reduction factor of $\text{BR}  (= \text{round}(N_b/L))$ by applying beam merging matrix $\mathbf{T}$ designed
by using a three-dimensional (3-D) beam merging network with 3-D convolutional kernels as shown in Figs.~\ref{fig:BSdualNet_FR} and \ref{fig:BSdualNet_FR_ND}. Jointly, the total REs for CSI-RS placement can be reduced by a factor of $\text{BR}\cdot\text{FR}$. Thus, the total number of pilot REs becomes $N_bK/(\text{BR}\cdot\text{FR})$.

The DL received signal vector $\mathbf{y}^{(i,k)}_{\text{DL}} \in \mathbb{C}^{L\times 1}$ at the $i$-th UE in the $k$-th RB can be expressed as  
\begin{equation}
\setlength{\abovedisplayskip}{4pt}
\setlength{\belowdisplayskip}{4pt}
\mathbf{y}^{(i,k)}_{\text{DL}}= \mathbf{S}^{(k)}_{\text{DL},L}\mathbf{T}^T\mathbf{h}^{(i,k)}_{\text{BS,DL}} + \mathbf{n}_{\text{DL}}^{(k)}\label{DL signal},
\end{equation}
where the superscript $(i,k)$ denotes the UE and RB indexes, respectively. Following
Section II, UE-$i$ 
estimates beam response vectors $\mathbf{g}^{(i,k)}_{\text{FB}}, k = 1, 1+FR,..., 1+(K-1)FR$ as a beam response matrix
\begin{gather*}
    \mathbf{G}_{\text{FB}}^{(i)}= \begin{bmatrix}\mathbf{g}^{(i,1)}_{\text{FB}}, \mathbf{g}^{(i,1+FR)}_{\text{FB}},\dots, \mathbf{g}^{(i,1+(K-1)FR)}_{\text{FB}} \end{bmatrix} \in \mathbb{C}^{L \times K/FR}
\end{gather*}
where the estimates $\mathbf{g}^{(i,k)}_{\text{FB}} = (\mathbf{S}^{(k)}_{\text{DL},L})^{-1}\mathbf{y}^{(i,k)}_{\text{DL}} \in \mathbb{C}^{L}$ are
based on pilots reduced by 
FR. 

\subsection{BSdualNet-FR}

For further reduction of UL feedback overhead, we compress the beam responses by implementing 
a frequency compression module (FCM) similar
to an autoencoder. The FCM consists of an encoder at UE and decoder at gNB for CSI compression and recovery, respectively. The encoder consists of four $3 \times 3$ circular convolutional layers with $16, 8, 4$ and $2$ channels. 
Subsequently, an FC layer with $\left \lceil{2LK/(\text{CR}\cdot\text{FR})}\right \rceil$ elements accounts for dimension reduction by a factor of $\text{CR}_\text{eff} = \text{BR}\cdot\text{FR}\cdot\text{CR}$ after reshaping. $\text{CR}_\text{eff}$ and $\text{CR}$ respectively denote the effective and feedback compression ratios. 
The FC layer output is sent to a quantization module which uses a trainable soft quantization function as proposed in \cite{CQNET} to generate feedback codewords.

At the gNB, the codewords from different UEs are forwarded into the decoder network of the FMC to recover their respective DL CSIs. The decoder first expands the dimension of the codewords to their original size of $2N_bK$. Reshaped into a size of $N_b \times K \times 2$, a codeword 
enters four $3 \times 3$ circular convolutional layers with with $16, 8, 4$ and 2 channels 
to generate the FCM output. Note that the dimensions in both frequency and beam domains are already the same as our target output in this stage. The FCM output serves as an initial DL CSI estimate $\widehat{\mathbf{H}}^{(i)}_{\text{BS,DL,ini}} \in \mathbb{C}^{N_b \times K}$ which is used to calculate the first loss
\begin{equation}
    \mbox{loss}_1 = \sum_{i=0}^{N-1}||\widehat{\mathbf{H}}^{(i)}_{\text{BS,DL,ini}} - \mathbf{H}^{(i)}_{\text{BS,DL}}||_2^2,
\end{equation}

\begin{equation}
\widehat{\mathbf{H}}^{(i)}_{\text{BS,DL,ini}} = f_\text{FMC,de}(f_\text{FMC,en}(\mathbf{G}^{(i)}_{\text{FB}})).
\end{equation}

Next, the combining network
refines the initial estimate with the 
help of UL CSI magnitudes. The combining network first split the magnitude 
and the phase of the initial estimate 
before sending the initial estimate magnitudes and the UL CSI magnitudes into five residual blocks which are constructed by a shortcut and four circular convolutional layers with $16, 8, 4, 2$ and $1$ channels and activation functions for magnitude refinement. 
From there, the refined magnitudes of DL CSI and their corresponding phases form the final output $\widehat{\mathbf{H}}^{(i)}_\text{BS,DL} \in \mathbb{C}^{N_b \times K}$ to determine the second loss function
\begin{equation}
   \mbox{loss}_2 = \sum_{i=0}^{N-1}||\widehat{\mathbf{H}}^{(i)}_{\text{BS,DL}} - \mathbf{H}^{(i)}_{\text{BS,DL}}||_2^2,
\end{equation}
\begin{equation}
    \mathbf{H}^{(i)}_{\text{BS,UL}} = \left[
\text{vec}(\mathbf{H}^{(i,1)}_{\text{BS,UL}}) \;
\text{vec}(\mathbf{H}^{(i,2)}_{\text{BS,UL}}) \;
\cdots\text{vec}(\mathbf{H}^{(i,K)}_{\text{BS,UL}})\right]
\end{equation}
\begin{equation}
\widehat{\mathbf{H}}^{(i)}_{\text{BS,DL}} = f_c(\widehat{\mathbf{H}}^{(i)}_{\text{BS,DL,ini}}, |\mathbf{H}^{(i)}_{\text{BS,UL}}|),
\end{equation}

The BSdualNet-FR is optimized by updating the network parameters $\Theta_\text{bm}$, $\Theta_\text{FMC,en}$, $\Theta_\text{FMC,de}$ and $\Theta_\text{c}$ of the non-linear 3-D beam merging, FMC encoder/decoder, and combining networks $f_\text{bm}$, $f_\text{FMC,en}$, $f_\text{FMC,de}$ and $f_\text{c}$:
\begin{gather*}
\setlength{\abovedisplayskip}{4pt}
\setlength{\belowdisplayskip}{4pt}
{\mathop{\arg\min}_{\Theta_\text{bm}, \Theta_\text{FMC,en}, \Theta_\text{FMC,de}, \Theta_\text{c}} \left\{\alpha \cdot 
\mbox{loss}_1 + (1-\alpha)\cdot \mbox{loss}_2
\right\}} \\ \label{loss_magnitude}
\end{gather*}
where hyperparameter $\alpha$ adjusts the weighting.


Note that the deep learning network contains many hyperbolic tangent activation functions and a soft quantization function which could lead to the gradient vanishing problem for 
parameters in those layers. To mitigate this problem, we suggest a two-stage training scheme for optimizing the proposed framework. In the first stage, we train the model by setting $\alpha = 1$ for $N_\text{first}$ epochs, freezing the combining network and focusing on finding the best beam merging matrix and encoding/decoding networks. In the second stage, we change $\alpha = 0.1$ and focus on refining the final estimates with the aid
of UL CSI magnitudes. Using the elbow method \cite{elbow_method}, we found that $N_\text{first} = 30$ is usually 
sufficient to obtain a good tradeoff.  

\section{Experimental Evaluations}
{
\subsection{Experiment Setup}
In our numerical test, we consider both indoor and outdoor cases. 
Using channel model software, 
we position a gNB of height equal to 20 m
at the center of a circular cell
with a radius of 30 m for indoor and 200 m for outdoor environment. 
We equip the gNB with a $8 \times 4 (N_H \times N_V)$ UPA for communication with 
single antenna UEs. UPA elements have half-wavelength uniform spacing.
The number of residual blocks in
the combining network is set to $N_B=5$ throughout. 

For our proposed model and other competing models, we set the number of epochs to $300$ and $1500$, respectively. We use batch size of $200$. For our model, we start with  learning rate of $0.001$ before switching to
$10^{-4}$ after the $100$-th epoch. 
Using the channel simulator, We generate several indoor and outdoor datasets, each containing 100,000 random channels. 
57,143 and 28,571 random channels are for training and validation. 
The remaining 14,286 channels are 
test data for performance evaluation. 
For both indoor and outdoor, we use 
the QuaDRiGa simulator \cite{QuaDriGa} 
using the scenario features given in \textit{3GPP TR 38.901 Indoor} and \textit{3GPP TR 38.901 UMa} at 5.1-GHz and 5.3-GHz, and 300 and 330 MHz of UL and DL with LOS paths, respectively. For both scenarios, $1024$ subcarriers with a $15$K-Hz spacing are considered for each subband. Here, we assume UEs are capable of perfect channel estimation. We set antenna type to \textit{omni}. We use normalized
MSE as
the performance metric
\begin{equation}
\frac{1}{ND}\sum^{D}_{d=1}\sum^{N}_{n=1}\norm{\widehat{\mathbf{H}}^{(i)}_{\text{BS,DL},d} - \mathbf{H}^{(i)}_{\text{BS,DL},d}}^2_\text{F} /\norm{\mathbf{H}^{(i)}_{\text{BS,DL},d}}^2_\text{F},\label{NMSE1}
\end{equation}
where the number $D$ and subscript $d$ denote the total number and index of channel realizations, respectively.

\subsection{Testing Different Numbers of Available REs}
We evaluate the performance of CSI recovery by adopting the proposed encoder-free CSI feedback frameworks, $\text{BSdualNet}_{0}$, $\text{BSdualNet}$ and $\text{BSdualNet-MN}$. To test the efficacy without considering quantization, we first compare $\text{BSdualNet}_{0}$ with two heuristic approaches (denoted as BS-UL and BS-DL) that recover DL CSIs according to $L$ beam responses where the beams are selected according to the UL and DL CSI magnitudes, respectively. Note that BS-UL should serve
as the lower bound of $\text{BSdualNet}_{0}$ since $\text{BSdualNet}_{0}$ is equivalent to refine the result of BS-UL with an additional combining network. 

Figs.~\ref{result_diffL} (a) and (b) provide the NMSE performance for different number of available REs $L$ in an RB for $\text{BSdualNet}_{0}$, BS-UL and BS-DL in both indoor and outdoor scenarios, respectively. The results show that $\text{BSdualNet}_{0}$ delivers better performance than BS-UL and also BS-DL in outdoor scenario owing to the high spatial correlation in beam domain. Because of the high angle spread induced by 
the more complex multi-path environment 
in indoor scenarios, the combining network in $\text{BSdualNet}_{0}$ only marginally improve the recovery performance.

Figs.~\ref{result_diffL_ad} (a) and (b) 
illustrate the NMSE performance for different number $L$ of REs within a RB for $\text{BSdualNet}_{0}$, $\text{BSdualNet}$ and $\text{BSdualNet-MN}$ for both indoor and outdoor channels, respectively. 
We can observe the benefits of the beam merging matrix $\mathbf{T}$ 
especially in outdoor cases. Furthermore, instead of using a convolution-layer based combining network, changing the combining
function as a minimum-norm solution yields a significant performance improvement in both indoor and outdoor scenarios. Since minimum-norm solution directly uses the beam merging matrix $\mathbf{T}$, it becomes more efficient to decouple the superposition of weighted beam responses by minimizing the MSE of DL CSIs.  

\begin{figure}
\centering
\resizebox{3.4in}{!}{
\includegraphics*{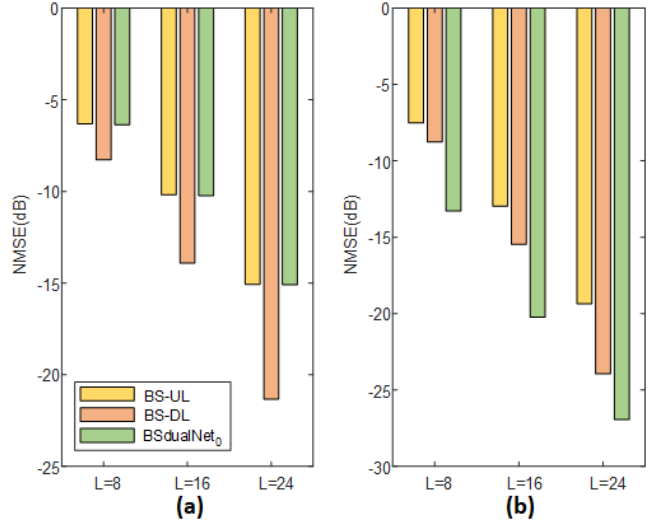}}
\caption{NMSE performance of BS-UL, BS-DL, and $\text{BSdualNet}_0$ for different TFRs $L$ in (a) indoor, (b) outdoor scenarios.\label{result_diffL}}
\end{figure}

\begin{figure}
\centering
\resizebox{3.4in}{!}{
\includegraphics*{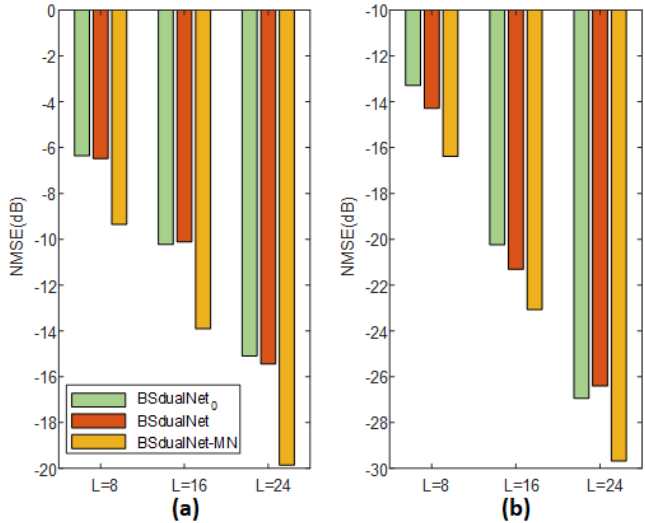}}
\caption{NMSE performance of $\text{BSdualNet}_0$, $\text{BSdualNet}$, and BSdualNet-MN for different TFRs $L$ in (a) indoor, (b) outdoor scenarios.\label{result_diffL_ad}}
\end{figure}

\subsection{Performance for Different Numbers of UEs}

Similar to our beam merging matrix $\mathbf{T}$,  measurement matrix in compressive sensing based frameworks \cite{ISTA, ISTANet} also functions 
to shrink the dimension of original data 
and derive a better representation for 
their sparsity that can be
easier to recover. 
To demonstrate the relative 
performance of the proposed frameworks, we also compare with two 
successful compressive approaches
ISTA \cite{ISTA} and ISTA-Net \cite{ISTANet}:

\begin{itemize}
    \item \textbf{Iterative Shrinkage-Thresholding Algorithm (ISTA)}: Its regularization parameter and maximum iteration number are set to $0.5$ and $3000$, respectively. 
    \item \textbf{ISTA-Net}: The phase and epoch numbers are set to $5$ and $1000$, respectively.
\end{itemize}

Figs.~\ref{result_L8_diffUE} (a) and (b) provide the NMSE performance comparison
for different numbers of UEs $N$ 
for $L=8$ REs in a RB for $\text{BSdualNet}$, $\text{BSdualNet-MN}$, ISTA and ISTA-Net 
and under indoor and outdoor scenarios, respectively. From the results, 
we observe the clear performance degradation for $\text{BSdualNet}$ and $\text{BSdualNet-MN}$ as UE number grows. This is intuitive since it is difficult
to find an optimum beam merging matrix 
for all active UEs. 
Fortunately, for most cases, the performance degradation tends to saturate after the UE number exceeds a certain number typically
less than $10$ for $\text{BSdualNet-MN}$. 

Our tests show that both $\text{BSdualNet}$ and $\text{BSdualNet-MN}$ deliver better
performance over ISTA and ISTA-Net under different UE numbers. Our heuristic insight
is that measurement matrix in ISTA and ISTA-Net is unknown at recovery whereas the beam merging matrix is designed
by the gNB and can be explicitly utilized
by the recovery decoders of $\text{BSdualNet}$ and $\text{BSdualNet-MN}$.

\begin{figure}
\centering
\resizebox{3.4in}{!}{
\includegraphics*{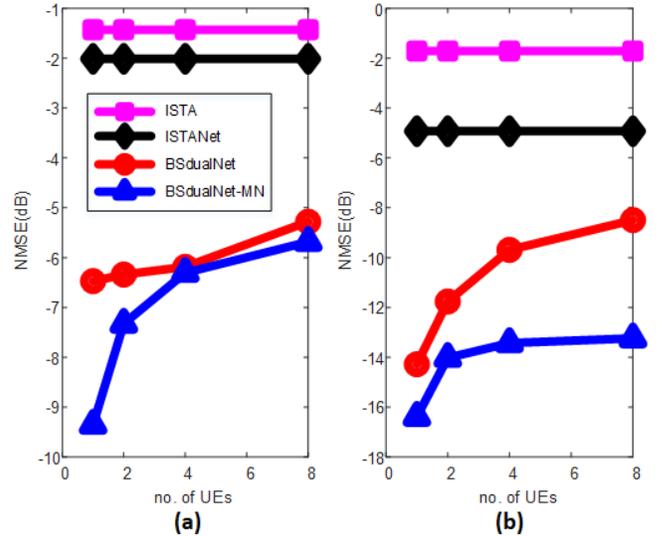}}
\caption{NMSE performance for different number of UEs $N$ in (a) indoor, (b) outdoor scenarios.\label{result_L8_diffUE}}
\end{figure}

\subsection{CSI-RS Configurations and Compression Ratios}
We consider a $5.76$ MHz subband (i.e., $32$ RBs each of bandwidth $180K$-Hz). Each codeword element uses 8 quantization bits. To comprehensively evaluate BSdualNet-FR, The two tables in Fig.~\ref{fig: sim_diff_cong_indoor} and Fig.~\ref{fig: sim_diff_cong_outdoor} provide the NMSE performance of BSdualNet-FR against different CSI-RS configurations and compression ratios in outdoor and indoor scenarios, respectively. We apply the
same background color on results with the same pilot and feedback overhead reduction ratios. 

Since outdoor channels generally 
exhibit stronger sparsity and larger delay spread respectively in beam and delay domains, we
observe a slight performance degradation
with BR increase as opposed to FR increase. Importantly, for $BR=4$, there is 
a clear performance loss
even when using the same pilot and feedback overhead reduction ratio. Despite the 
channel sparsity, with
the use of half-wavelength antenna spacing
(i.e., Nyquist sampling in spatial domain), the overly aggressive compression in beam domain cause too much information loss
to recovery at the gNB. 
For indoor channels, we observe
a slight performance degradation when increasing FR instead of BR because of 
larger angular and shorter delay spread
of indoor CSI.
\begin{figure}
\centering
\resizebox{3.2in}{!}{
\includegraphics*{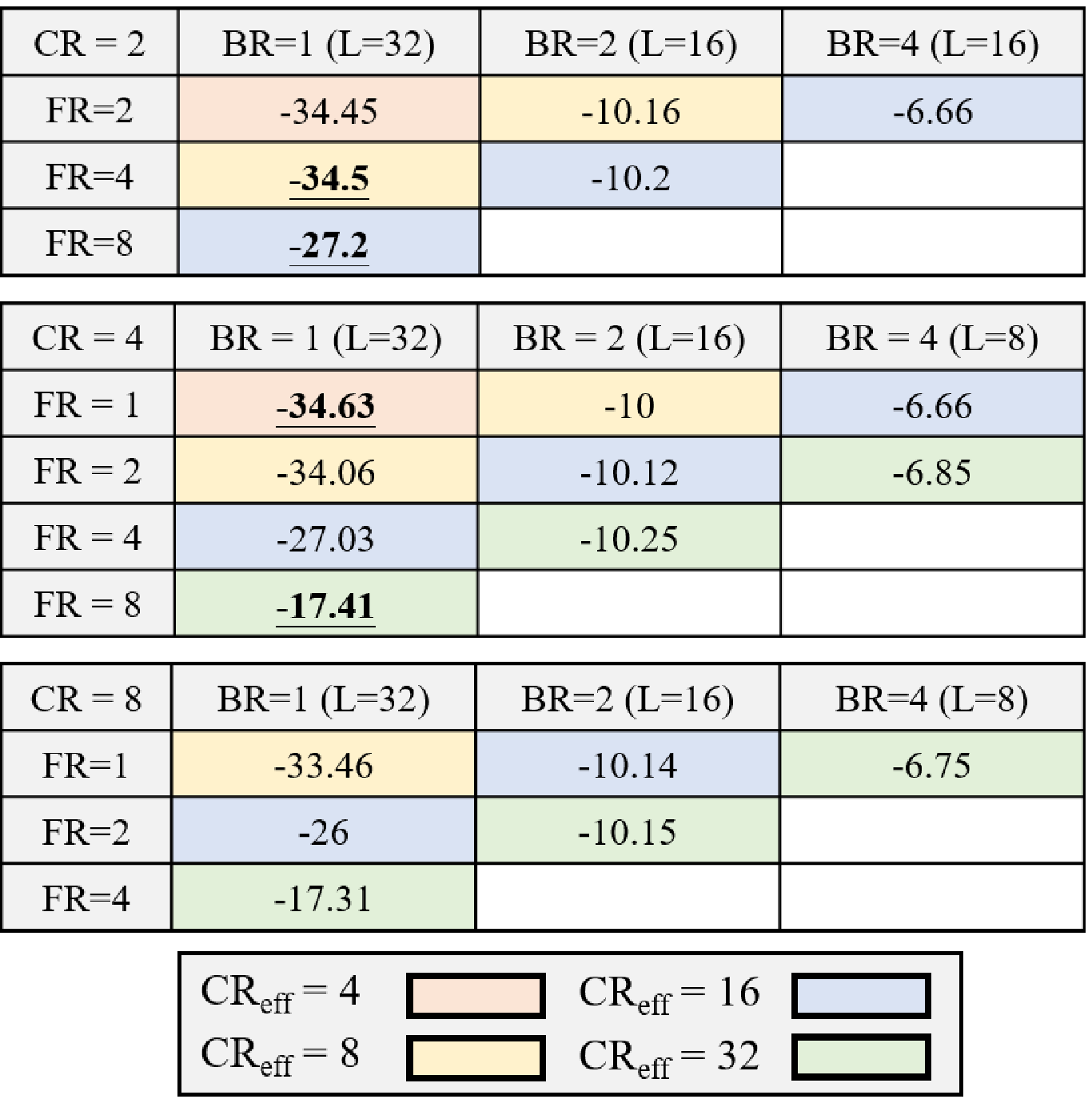}}
\caption{NMSE performance of BSdualNet-FR for different CSI-RS placement configurations in indoor scenarios. (The results with the same effective compression ratio are denoted as the same color. The best performance at the same effective compression ratio is denoted by bold fonts with underline.)\label{fig: sim_diff_cong_indoor}}
\end{figure}
\begin{figure}
\centering
\resizebox{3.2in}{!}{
\includegraphics*{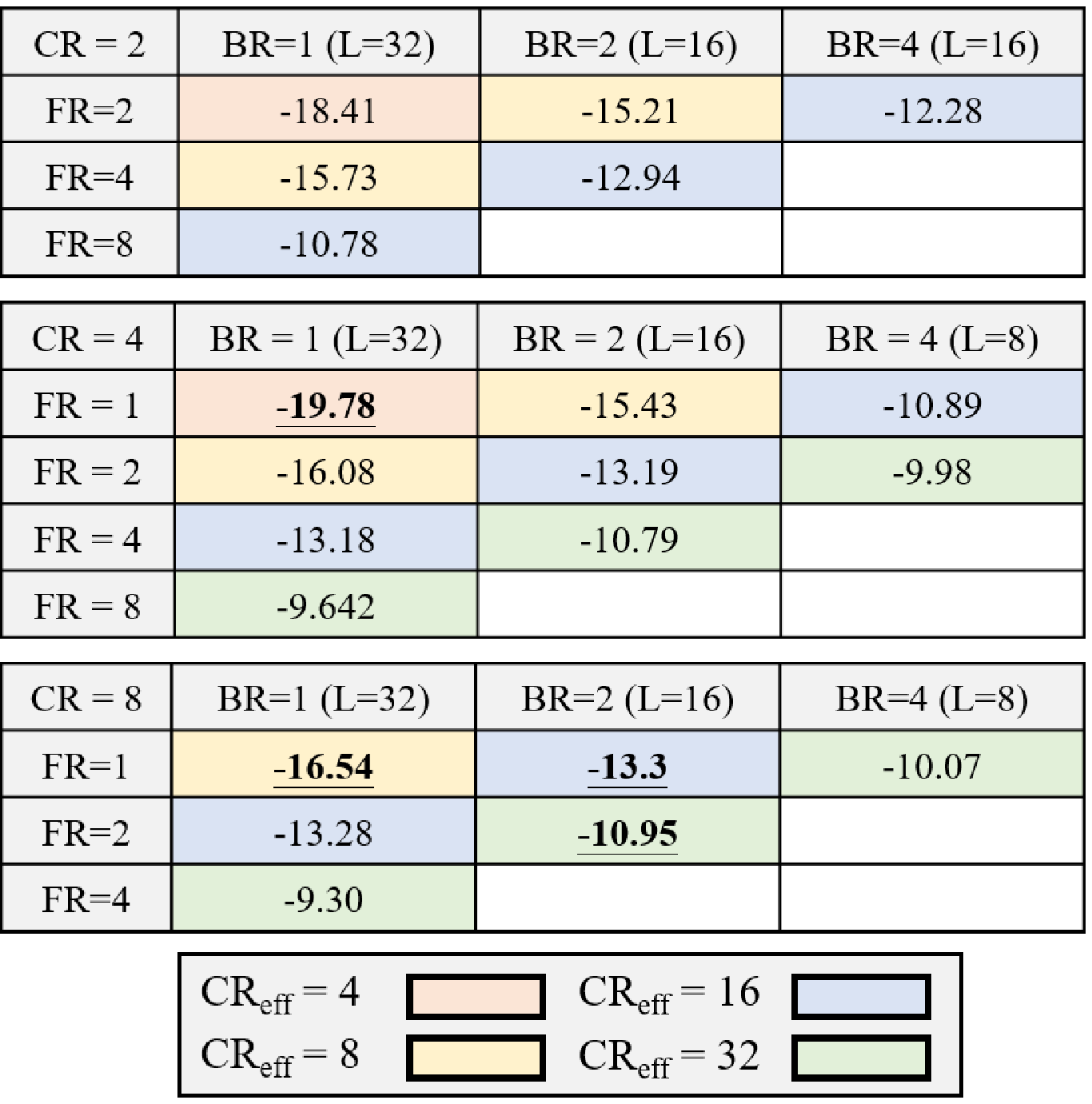}}
\caption{NMSE performance of BSdualNet-FR for different CSI-RS placement configurations in outdoor scenarios. (The results with the same effective compression ratio are denoted as the same color. The best performance at the same effective compression ratio is denoted by bold fonts with underline.)\label{fig: sim_diff_cong_outdoor}}
\end{figure}

\subsection{Effective Compression Ratio $\text{CR}_\text{eff}$}
As benchmarks, we also compare BSdualNet-FR with CsiNet-Pro \cite{MarkovNet} and another successful method DualNet-MP \cite{DualNet-MP}. The newly proposed
DualNet-MP also exploits FDD reciprocity 
by incorporating UL CSI magnitude as side information at CSI decoder of gNB. Table~\ref{table: Diff_CReff} presents the three way comparison of NMSE for CsiNet-Pro, DualNet-MP, and BSdualNet-FR under different values of effective compression ratio 
$\text{CR}_\text{eff}$ in indoor and outdoor cases. Benefiting from the UL CSI magnitudes, both BSdualNet-FR and DualNet-MP can outperform CsiNet-Pro in most cases. Interesting, better utilization of UL CSI by BSdualNet-FR provides better performance than DualNet-MP. Although the performance gain becomes less impressive for higher $\text{CR}_\text{eff}$, the additional
benefit of the BSdualNet-FR framework is
the reduction of REs for DL CSI-RS by a factor of $\text{BR}\cdot\text{FR}$ 
that allows gNB to 
reconfigure the CSI-RS placement to
enhance the DL spectrum efficiency. 

\begin{table*}[ht]
\centering
\caption{NMSE performance of different CSI feedback frameworks at different $\text{CR}_\text{eff}$.\label{table: Diff_CReff}}
\begin{tabular}{|c|c|c|c|c|c|c|} 
\hline
      & \multicolumn{2}{c|}{CsiNet-Pro} & \multicolumn{2}{c|}{DualNet-MP} & \multicolumn{2}{c|}{BSdualNet-FR}                                                                                                         \\ 
\hline
$\text{CR}_\text{eff}$ & Indoor   & Outdoor              & Indoor   & Outdoor              & Indoor                                                              & Outdoor                                                             \\ 
\hline
4     & -24.2 & -13            & -27.3 & -19.1             & \begin{tabular}[c]{@{}c@{}}\textbf{-34.6}\\(FR = 1, BR = 1)\end{tabular}  & \begin{tabular}[c]{@{}c@{}}\textbf{-19.8}\\(FR = 1, BR = 1)\end{tabular}  \\ 
\hline
8     & -20.8 & -12.5            & -20.9 & -16.4             & \begin{tabular}[c]{@{}c@{}}\textbf{-34.5}\\(FR = 4, BR = 1)\end{tabular}  & \begin{tabular}[c]{@{}c@{}}\textbf{-16.5}\\(FR = 1, BR = 1)\end{tabular}  \\ 
\hline
16    & -14.4 & -11.8              & -20.2 & -13.3             & \begin{tabular}[c]{@{}c@{}}\textbf{-27.2}\\(FR = 8, BR = 1)\end{tabular}  & \begin{tabular}[c]{@{}c@{}}\textbf{-13.3}\\(FR = 1, BR = 2)\end{tabular}  \\ 
\hline
32    & -13.2  & -8.6          & -16.8 & -11             & \begin{tabular}[c]{@{}c@{}}\textbf{-17.4}\\(FR = 8, BR = 1)\end{tabular} & \begin{tabular}[c]{@{}c@{}}~\textbf{-11}\\(FR = 2, BR = 2)\end{tabular}  \\
\hline
\end{tabular}
\end{table*}

\subsection{Complexity: FLOPs and Parameters}

\begin{table*}[ht]
\centering
\caption{Comparison of parameters and FLOPs at encoder.\label{Complexity_Diff_Encoding}}
\begin{tabular}{|c|c|c|c|c|c|c|} 
\hline
  & \multicolumn{2}{c|}{CsiNet-Pro} & \multicolumn{2}{c|}{DualNet-MP} & \multicolumn{2}{c|}{BSdualNet-FR}            \\ 
\hline
$\text{CR}_\text{eff}$ & Parameters & FLOPs          & Parameters  & FLOPs             & Parameters  & FLOPs                          \\ 
\hline
4      & 1M   & 4.23M    & 0.54M & 4.2M        & 1M/(FR*BR) 
 & (2.1 + 2.1/(FR*BR))M        \\ 
\hline
8      & 534K & 2.12M    & 280K  & 2.2M        & 534K/(FR*BR) 
 & (1.1 + 1/(FR*BR))M        \\ 
\hline
16     & 272K & 1.08M    & 140K  & 1.1M        & 272K/(FR*BR) 
 & (0.55 + 0.5/(FR*BR))M   \\ 
\hline
32     & 140K & 0.56M    & 82K   & 0.6M        & 140K/(FR*BR) 
 & (0.27 + 0.26/(FR*BR))M  \\
\hline
\end{tabular}
\end{table*}

Most UEs have stronger memory, computation, and power constraints. The system
design favors light-weight and simpler encoders for
deployment at UEs. In comparison with
the baseline CsiNet Pro, Table~\ref{Complexity_Diff_Encoding} shows dimension reduction in frequency and beam domains and smaller input size of our encoder/decoder architecture.
BSdualNet-FR provides significant 
reduction in terms of FLOPs and 
the number of model parameters. 
Similarly, if
the total reduction factor $\text{FR}\cdot \text{BR}\ge 2$}, 
BSdualNet-FR shows lower complexity than DualNet-MP.


\section{Conclusions}
This work presents a new deep learning framework for CSI estimation 
in massive MIMO downlink. 
Leveraging UL CSI estimate to reduce its CSI-RS resources, the gNB designs a beam merging matrix based on UL channel magnitude
information to transform DL CSI observation
at UEs into a lower dimensional representation that is easier for
feedback and recovery. We further
develop an efficient minimum-norm CSI recovery network to improve recovery accuracy. Our new framework does not
deploy training deep learning models at
UEs, thereby lowering UE complexity
and power consumption. 
We achieve further reduction
of DL CSI training and feedback overhead, 
by introducing a reconfigurable CSI-RS placement. 
Test results demonstrate significant improvement of CSI recovery accuracy 
and reduction of both DL CSI training and UL feedback overheads. 

\section*{Appendix}
\setcounter{equation}{0}
\setcounter{subsection}{0}
\renewcommand{\theequation}{A.\arabic{equation}}
\renewcommand{\thesubsection}{A.\arabic{subsection}}
\subsection*{Proof of Eq. (\ref{eq: I}):}
For an $L\times N_b$ merging matrix $\mathbf{T}$ with $L< N_b$, 
we have an underdetermined linear problem $\mathbf{y} = \mathbf{T}\mathbf{x}$. The minimum norm solution is simply
\begin{equation}
\begin{aligned}
    \mathbf{x}_\text{mn} & = \mathbf{T}^H(\mathbf{T}\mathbf{T}^H)^{-1}\mathbf{T}\mathbf{x},\\
\end{aligned}
\end{equation}
Based on singular value decomposition of  $\mathbf{T}$ by
\begin{equation}
    \mathbf{T} = \mathbf{U}\left[\begin{array}{cc}
    \mathbf{\Sigma} & \mathbf{0}\end{array}\right]\mathbf{V}^H, \label{eq:SVD}
\end{equation}
where $\mathbf{U}$ and $ \mathbf{V}$ 
respectively are left and right singular matrices corresponding
to the $L\times L$ diagonal $\mathbf{\Sigma}$ of nonzero singular
values. Let $ \mathbf{V}=[
\mathbf{v}_1\; 
\mathbf{v}_2\; 
\cdots
\mathbf{v}_{N_b}]
$ denote the corresponding right
singular vectors. It is clear that
\begin{align}
\mathbf{T}^H(\mathbf{T}\mathbf{T}^H)^{-1}\mathbf{T}= \mathbf{V}\left[\begin{array}{cc}\mathbf{I}_{L\times L} & \mathbf{0}\\
\mathbf{0}&\mathbf{0}\end{array}\right]
\mathbf{V}^H=\sum_{i=1}^L \mathbf{v}_i
\mathbf{v}_i^H
\end{align}
Define a matrix $\widetilde{\mathbf{I}}=
\sum_{i=1}^L \mathbf{v}_i
\mathbf{v}_i^H$. 
The minimum-norm solution is simply
\begin{equation}
    \mathbf{x}_\text{mn} = \sum_{i=1}^{L}\mathbf{v}_i\mathbf{v}_i^H\mathbf{x} = \widetilde{\mathbf{I}}\cdot \mathbf{x}. 
\end{equation}
Since the singular vectors $\{\mathbf{v}_i\}$ are orthonormal, i.e.,
$\mathbf{v}_i^H\mathbf{v}_i=1$, 
it is
clear that
\begin{align}
\mbox{Trace}\{\widetilde{\mathbf{I}}\}
&=\sum_{i=1}^L  \mbox{Trace}\{ \mathbf{v}_i
\mathbf{v}_i^H\}\nonumber \\
&=\sum_{i=1}^L  \mbox{Trace}\{ \mathbf{v}_i^H
\mathbf{v}_i\}\label{EqA5}\\
&=\sum_{i=1}^L 1=L
\end{align}
in which the equality of Eq.~(\ref{EqA5}) 
holds because $\mbox{Trace}
\{\mathbf{A}\mathbf{B}\}=\mbox{Trace}
\{\mathbf{B}\mathbf{A}\}$.

\section{Acknowledgement}
The authors would like to acknowledge Mason del Rosario for his useful discussions which helped the authors better understand of pilot placement and channel truncation.

\bibliography{references.bib}
\bibliographystyle{IEEEtran}
\end{document}